\documentclass[runningheads]{llncs}
\usepackage{graphicx}
\usepackage{epsfig}
\usepackage{graphics}
\usepackage{array}
\usepackage{comment}
\date{}
\title{A Note on Contractible Edges in Chordal graphs}
\titlerunning{Contractible Edges in Chordal graphs}
\author{N.S.Narayanaswamy  \and N.Sadagopan  \and Apoorve Dubey   } 
\institute{Department of Computer Science and Engineering, Indian Institute of Technology, Chennai-600036, India. \\
\email{\{\{swamy,sadagopu\}@cse.iitm.ac.in,apoorvedubey82@gmail.com\}}}
\begin{document}
\maketitle
\begin{abstract}
Contraction of an edge merges its end points into a new vertex which is adjacent to each neighbor of the end points of the edge.  An edge in a $k$-connected graph is {\em contractible} if its contraction does not result in a graph of lower connectivity.  We  characterize contractible edges in chordal graphs using properties of tree decompositions with respect to minimal vertex separators.
\end{abstract}
\section{Introduction}
Chordal graphs are also known as triangulated graphs \cite{rose1} and have applications in the study of linear sparse systems, scheduling and relational database systems.  In this paper, we focus on $k$-connected chordal graphs.  We study the impact of contraction on connectivity in $k$-connected chordal graphs.
In a graph $G$, contraction of an edge $e$ with endpoints $u,v$ is the replacement of $u$ and $v$ with a single vertex $z$.  In the resulting graph, the edges incident on $u$ and $v$ are incident on $z$.  Edge contraction and in general clique contraction plays a significant role in the proof of the Perfect Graph Theorem, see  \cite{golu}.     Edge contraction also plays a significant role in min-cut algorithms by using the basic property that the contraction of an edge can only increase the size of the min-cut.   The basic idea exploited in randomized algorithms for min-cut is that contracting a randomly chosen edge does not increase the size of the min-cut \cite{karg}.  This leads to expected polynomial time algorithms for min-cut, and these algorithms are fundamentally different from the classical max-flow based techniques. 
\subsection{Past Results on Contractible Edges}
As with many a problem in Graph Theory, the study of contractible edges was initiated by Tutte in \cite{tutte} where a constructive characterization of  3-connected graphs was presented.   One consequence of this characterization was that in any 3-connected graph with at least five vertices, there is at least one contractible edge.   In the work by Saito et. al \cite{saito}, this lower bound was improved to $\frac{|V(G)|}{2}$, and the structure of graphs that have exactly so many contractible edges is studied.  For $k$-connected graph with $k\geq 4$,  it is still ongoing research to find necessary and sufficient conditions for the presence of contractible edges.  For example, Thomassen \cite{thomassen} has shown that there is a contractible edge in  triangle-free $k$-connected graphs in which the minimum degree is more than $\frac{3k-3}{2}$.   Kriesell's survey of contractible edges  \cite{matthias} is an excellent source for many results in this area, and is also the motivation point of our work.
\subsection{Definitions}
We have, to a large extent, followed the notation and definitions as in the Graph Theory text by West\cite{west}. 
\noindent
Let $G =(V,E)$ be an undirected non weighted graph where $V(G)$ is the set of vertices and $E(G) \subseteq \{\{u,v\} | u,v \in V(G)$, $u \not= v \}$.  Order of $G$ and size of $G$ are $|V(G)|$ and $|E(G)|$, respectively.  The {\em neighborhood} of a vertex $v$ in a graph $G$ is the set $\{u | \{u,v\} \in E(G)\}$ and is denoted by $N_G(v)$.  A separating set or cut set of a graph $G$ is a set $S \subseteq V(G)$ such that the induced subgraph, denoted by $G-S$, on the vertex set $V(G) \setminus S$ has more than one connected component.  The vertex connectivity of a graph $G$, written  $\kappa(G)$, is the minimum cardinality of a vertex set $S$ such that $G-S$ is disconnected or has only one vertex.  $\gamma_G$ is the set of all minimum order cut sets. We let $G.e$ denote the graph obtained by contracting an edge $e=\{u,v\}$ in $G$ such that $V(G.e)=V(G)\setminus\{u,v\}\cup\{z_{uv}\}$ and $E(G.e)=\{ \{z_{uv},x\} \vert \{u,x\}$ or $\{v,x\} \in E(G)\} \bigcup \{\{x,y\} \vert x \not= u, y \not= v$ in $E(G) \}$.  An edge $e \in E(G)$ is contractible if  the connectivity of $G.e$ is same as the connectivity of $G$.  $E_c(G)$ denotes the set of contractible edges in $G$.   A $k$-connected graph $G$ is said to be contraction critical, if for each edge $e$, connectivity of $G.e$ is smaller than the connectivity of $G$.
The following lemma relates cut sets and contractible edges \cite{matthias}.
\begin{lemma}
\label{contractionlemma}
An edge $e=\{u,v\}$ of $G$ is non contractible if and only if  there is a minimum cut set $T \in \gamma_G$ such that $u \in T $and $ v \in T$.
\end{lemma}
A tree decomposition of a graph $G=(V,E)$ is a tree $T$, where each node $x$ has a label $l(x) \subseteq V(G)$ such that:
\begin{itemize}
\item{$\displaystyle \bigcup_{x \in V(T)} l(x)=V(G)$.(We say that "all vertices are covered.")}
\item{For any edge $\{v,w\} \in E(G)$, there exists a node $x$ in $T$ such that $v,w \in l(x)$.(We say that "all edges are covered.")}
\item{For any $v \in V(G)$, the set of all nodes of $T$ whose label contains $v$ form a connected subtree in $T$.(We call this the "connectivity condition")}
\end{itemize}
{\bf Chordal Graph Preliminaries} \\
A {\em chord} of a cycle $C$ is an edge not in $C$ whose endpoints lie in $C$.  A {\em chordless cycle} in $G$ is a cycle of length at least $4$ in $G$ that has no chord.  A graph $G$ is {\em chordal} if it is simple and has no chordless cycle.  We can represent a chordal graph $G$ using a tree decomposition $T$ as follows: for each vertex $x \in V(T)$ the associated label $l(x) \subseteq V(G)$ induces a maximal clique in $G$, and for each $v \in V(G)$, $T_{v}$,  the subgraph of $T$ induced by the set $\{x \in V(T) | v \in l(x)\}$, is a tree.   We use $M$ to denote the set of minimal vertex separators of $G$, and the graph to which the symbol $M$ applies is always clear from the context.  A stable (or independent) set is a set of pairwise nonadjacent vertices of the graph $G$.   A {\em split} graph $G$ is a graph with two partitions, a stable set $I$ and a clique $K$ such that $V(G)=I \bigcup K$. $E(G) \subseteq \{\{u,v\} | u \in I, v \in K \}$. For a chordal graph $G$ and its tree decomposition $T$, a minimal vertex separator $S$, and an edge $e \in E(G)$, we consider fixed tree decompositions of $G \setminus S$ and $G.e$, denoted by $T \setminus S$ and $T.e$, respectively.  
$T \setminus S$ and $T.e$ are defined as follows: The vertex set of both $T \setminus S$ and $T.e$ are same as the vertex set of $T$. The removal of $S$ and the contraction of $e$ only changes the labels associated with the vertices.  In $T \setminus S$, for each $x \in V(T \setminus S)$, we write $l(x) = l(x) \setminus S$, if $S \cap l(x) \not = \phi$, otherwise $l(x)$ is the same set as in $T$.  In $T.e$, for each $x \in V(T.e)$, $l(x) =  l(x) \setminus \{u,v\} \cup \{z_{uv}\}$, if $l(x) \cap \{u,v\} \not = \phi$.  Otherwise, $l(x)$ is the same set as in $T$.  Clearly, $T \setminus S$ and $T.e$ are tree decompositions of $G \setminus S$ and $G.e$, respectively.
\section{The Structure of Contractible edges in $k$-connected Chordal Graphs}
We first prove a theorem which characterizes the set of minimal vertex separators of a chordal graph.  This result is used subsequently to prove our characterisation of contractible edges in chordal graphs.  
\begin{lemma}
\label{charlemma}
Let $G$ be a chordal graph and $T$ be its tree decomposition.  $G$ is connected iff for each edge $\{x,y\} \in E(T)$, $l(x) \cap l(y)  \not= \phi$.
\end{lemma}
\begin{proof}
{\em Necessity:} If $G$ is connected then we need to show that for each edge $\{x,y\} \in E(T)$, $l(x) \cap l(y) \not= \phi$.  We prove this by contradiction.  Suppose there exists an edge $\{x,y\} \in E(T)$ and  $l(x) \cap l(y) = \phi$.  Consider the two components $C_1$ and $C_2$ obtained by removing the edge $\{x,y\}$. Assume that $x \in C_1$ and $y \in C_2$.  Let $A=\displaystyle \bigcup_{z \in C_1} l(z)$,  $B=\displaystyle \bigcup_{z \in C_2} l(z)$.  Since $T$ is a tree decomposition and $l(x) \cap l(y) = \phi$, it follows that $A \cap B= \phi$.  Further, each edge $e \in E(G)$ is contained in the graph induced by $A$ or $B$ but not both. Hence $G$ is disconnected.  However, by our hypothesis $G$ is connected.  Hence our assumption is wrong.  Therefore, if $G$ is connected then for each edge $\{x,y\} \in E(T)$, $l(x) \cap l(y) \not= \phi$. \\
{\em Sufficiency:}  Given that for each edge $\{x,y\} \in E(T)$, $l(x) \cap l(y) \not= \phi$, we now show that $G$ is connected.  We show that $\forall u, \forall v \in V(G), u\not=v$,  there exists a path between $u$ and $v$ in $G$. Let $x,y$ be any two vertices in $V(T)$ such that $u \in l(x)$ and $v \in l(y)$.  Consider the path $x=z_1,z_2,...,z_j=y$ in the tree $T$.  Further, $l(z_i) \cap l(z_{i+1})$ is non empty in $T$.  This implies that there exists a vertex $r_i \in l(z_i) \cap l(z_{i+1})$.  Hence the sequence of edges $\{u,r_1\}\{r_1,r_2\}... \{r_{j-1},v\}$ is a $uv$ path in $G$.  The reason this is true in $G$ is because $G$ is a chordal graph and label of each node in $T$ is a maximal clique.  Therefore $u$ and $v$ are connected in $G$.  Hence $G$ is connected.  
\end{proof}
{\bf Note:} For a simple graph 
 $G$ and any tree decomposition $T$, if $G$ is connected then for each edge $\{x,y\} \in E(T)$, $l(x) \cap l(y)  \not= \phi$.

\begin{theorem}
\label{chordaltreedecomp}
Let $G$ be a $k$-connected chordal graph and let $T$ be its tree decomposition.  Let $M' = \{X|X = l(x) \cap l(y) \mbox{ where } \{x,y\} \in E(T)\}$ and $M''=\{Y | Y \in M' $ and for all $Z \in M',  Z \not \subset Y \}$. $M = M''$.  In other words, $M''$ is the set of minimal vertex separators of $G$.
\end{theorem}
\begin{proof}
{\bf $M'' \subseteq M$}:\\
Let $S \in M''$.  Clearly, for some $\{x,y\} \in E(T), S=l(x) \cap l(y)$. By definition, $T \setminus S$ is a tree decomposition of $G \setminus S$ and $l(x) \cap l(y) = \phi$. From Lemma \ref{charlemma} it follows that $G \setminus S$ is disconnected. Hence $S$ is a vertex separator.  Further, since $S \in M''$, it follows that there is no  $S' \subset S$ such that $G \setminus S'$ is disconnected.  Therefore, $S$ is a minimal vertex separator. 

\noindent
{\bf $M \subseteq M''$}\\
Let $S$ be a minimal vertex separator(MVS). We now show that $S \in M''$. We now argue that there exist distinct $x,y \in V(T)$
such that $\{x,y\} \in E(T) $ and $l(x) \cap l(y) = S$.  Since $S$ is a MVS, $G \setminus S$ is disconnected.  Since $G \setminus S$ is disconnected, in $T \setminus S$ there exists a pair of vertices $x$ and $y$ such that $l(x) \cap l(y) = \phi$.  We now claim that in $T$, $S=l(x) \cap l(y)$.  If $l(x) \cap l(y) \subset S$, this implies that $S$ is not a MVS.  However we are given the fact that $S$ is a MVS.  Therefore, $S=l(x) \cap l(y)$ and consequently $S \in M'$.  Since $S$ is a MVS it follows that $S \in M''$.  Hence the proof.
\end{proof}
Using the above characterization based on minimal vertex separators and tree decomposition of chordal graphs, we present the following theorem on the structure of contractible edges.
\begin{theorem}
\label{chordallemma}
Let $G$ be a $k$-connected chordal graph with $|V(G)| \geq (k+2)$ . An edge $e =\{u,v\} \in E(G)$ is contractible iff one of the following holds\\
(i) $e$ is in a unique maximal clique in $G$ \\
(ii) For $x,y \in V(T)$, $\{u,v\} \subset l(x) \cap l(y) $ and $\{x,y\} \in E(T)$, $|l(x) \cap l(y)| > k$.
\end{theorem}
\begin{proof}
{\em Necessity:} \\
$(i)$: Given that $e$ is contractible implies that $G.e$ is $k$-connected. If $e$ is in a unique maximal clique in $G$, then we are done.  In the case when $e$ is not in a unique maximal clique, let $e \in l(x) \cap l(y)$ for some $\{x,y\} \in E(T)$.  We now show that
$|l(x) \cap l(y)| > k$.  We prove this claim by contradiction.  Let us assume that $|l(x) \cap l(y)| \leq k$.  On contraction of $e$ in $G$, the tree decomposition of $G.e$ is $T.e$.  In $T.e$, the $|l(x) \cap l(y)| \leq k-1$.  From lemma \ref{charlemma} it follows that that $l(x) \cap l(y)$ is a vertex separator of $G.e$, and since $|l(x) \cap l(y)| \leq k-1$, it follows that $G.e$ is $k-1$-connected.  This is a contradiction to 
the hypothesis that $G.e$ is $k$-connected.  Therefore, our assumption that $|l(x) \cap l(y)| \leq k$ is wrong.  It follows that
$|l(x) \cap l(y)| > k$.  \\
{\em Sufficiency:}  First, we consider the case when $e$ is in a unique maximal clique and show that $e$ is contractible.  If $e$ is in a unique maximal clique in $G$  implies that $e$ is contained in the label of a unique node in $T$.  Therefore, for each $x,y \in T$, $|l(x) \cap l(y)|$ remains unchanged in $T.e$.  From theorem \ref{chordaltreedecomp} the connectivity of $G.e$ is at least as much as the connectivity of $G$.  Therefore, $e$ is contractible.   In the case when
 $|l(x) \cap l(y)| > k$ for all $\{x,y\} \in E(T)$, after contracting $e$, in $T.e$ $|l(x) \cap l(y)|$ is at least $k$ and hence the connectivity of $G.e$ is at least $k$, by theorem \ref{chordaltreedecomp}.  Hence $G.e$ is $k$-connected.  Therefore, $e$ is contractible in $G$.
\end{proof}

\noindent
As a consequence of this lemma, it follows that each edge incident on a simplicial vertex in a $k$-connected chordal graph is contractible.  Therefore, a $k$-connected chordal graph has at least $2k$ contractible edges.  We now apply the main lemma to understand contractible edges in split graphs.  Let $G$ be a non regular split graph.  An edge $e=\{u,v\}$ such that $u \in K$ and $v \in I$ is contractible. Clearly such an edge $e$ is in a unique maximal clique in $G$.  By theorem \ref{chordallemma} $e$ is contractible.
For the case when $G$ is a regular $k$-connected split graph with at least $k+2$ vertices, it follows that $G$ is contraction critical, that is none of the edges of $G$ are contractible.  The reason is that, given that $G$ is regular implies that there is exactly one vertex in $I$. Thus the resulting graph is a complete graph and 
each edge in every complete graph is non contractible. Therefore, $G$ is contraction critical. 
\bibliographystyle{splncs}
\bibliography{version1}
\end{document}